\documentstyle[twoside,12pt]{article}
\normalsize
\def\greaterthansquiggle{\raise.3ex\hbox{$>$\kern-.75em\lower1ex\hbox{$\sim$}}}
\def\lessthansquiggle{\raise.3ex\hbox{$<$\kern-.75em\lower1ex\hbox{$\sim$}}}
\newcommand{\bdi}{\begin{displaymath}}
\newcommand{\edi}{\end{displaymath}}
\newcommand{\bfi}{\begin{figure}}
\newcommand{\efi}{\end{figure}}

\newcommand{\beq}{\begin{equation}}
\newcommand{\eeq}{\end{equation}}

\newcommand{\gaf}{\gamma_{5}}

\newcommand{\beqa}{\begin{eqnarray}}
\newcommand{\eeqa}{\end{eqnarray}}
\newcommand{\no}{\nonumber}

\newcommand{\ra}{\rightarrow}

\newcommand{\wt}{\widetilde}

\def\au{{\setbox0=\hbox{\lower1.36775ex%
\hbox{''}\kern-.05em}\dp0=.36775ex\hskip0pt\box0}}
\def\ao{{}\kern-.10em\hbox{``}}
\newcommand{\dsla}{\partial\hspace{-6pt} /  }  
\newcommand{\Asla}{A\hspace{-6.5pt}  /  }

\begin{document}
\bibliographystyle{plain}

\begin{titlepage}
\begin{flushleft} 
FSUJ TPI 16/96
\end{flushleft}
\begin{flushright}
October, 1996
\end{flushright}

\vspace{1cm}
\begin{center}
{\Large \bf The three-boson bound state in the massive Schwinger model}\\[1cm]
C. Adam* \\
Friedrich-Schiller-Universit\"at Jena, Theoretisch-Physikalisches Institut \\
Max-Wien Platz 1, D-07743 Jena, Germany
\vfill
{\bf Abstract} \\
\end{center}

We derive the (matrix-valued) Feynman rules of mass perturbation theory of the
massive Schwinger model for non-zero vacuum angle $\theta$. Further, we discuss
the properties of the three-boson bound state and compute -- by a partial 
resummation of the mass perturbation series -- its mass and its partial decay
widths. 

\vfill

$^*)${\footnotesize permanent address: Inst. f. theoret. Physik d. Uni Wien \\
Boltzmanngasse 5, 1090 Wien, Austria \\
email address: adam@pap.univie.ac.at}
\end{titlepage}

\section{Introduction}

The massive Schwinger model is two-dimensional QED with one massive fermion. In
this model some nontrivial field-theoretic features occur. E.g. instanton-like
gauge field configurations are present in the model and, therefore, a
$\theta$-vacuum has to be introduced as a new, physical vacuum 
(\cite{CJS,Co1}). Further,
confinement is realized in this model and there are no fermions in the physical
spectrum (\cite{AAR, GKMS,CONF}). 
The fundamental physical particle of the theory is a massive, scalar
boson (the Schwinger boson with mass $\mu$). This boson is interacting and may
form bound states (\cite{Co1,BOUND,GBOUND}) and undergo scattering
(\cite{SCAT}). A proper way to discuss the model is mass perturbation theory
(i.e. perturbation in the fermion mass term), as it preserves the nontrivial
structure of the model (like e.g. $\theta$-vacuum) and is performed in terms of
physical (bosonic) variables only (\cite{Co1,FS1,MSSM,SMASS}). 
This mass perturbation theory uses the known
exact solution of the massless Schwinger model as a starting point (in the
massless model the Schwinger boson is a free scalar particle with mass $\mu_0
=\frac{e}{\sqrt{\pi}}$; \cite{AAR}, \cite{Sc1} -- \cite{Adam}).

In this paper we first derive the Feynman rules of the mass perturbation theory
for general vacuum angle $\theta$ (which slightly complicates them as they
acquire a matrix structure). We then discuss the three-boson bound state and
show how, by a resummation of the perturbation series, we may compute the mass
$M_3$ of the three-boson bound state and its partial decay widths for the
decays into two Schwinger bosons and into one Schwinger boson and one two-boson
bound state.

The mass perturbation theory that is used throughout this paper is based on the
Euclidean path integral formalism and, therefore, we have to take into account
our specific Euclidean conventions (see e.g. \cite{ABH}).

\input psbox.tex
\let\fillinggrid=\relax
\section{Mass perturbation theory}

First let us shortly review the mass perturbation theory. By simply expanding
the mass term, the vacuum functional and VEVs of the massive model may be
traced back to space-time integrations of VEVs of the massless model. E.g. the
vacuum functional is
\bdi
Z(m,\theta )= \sum_{k=-\infty}^\infty e^{ik\theta}N\int D\bar\Psi D\Psi
DA_\mu^k \cdot
\edi
\beq
\cdot \sum_{n=0}^\infty \frac{m^n}{n!}\prod_{i=1}^n \int dx_i \bar\Psi
(x_i)\Psi (x_i)\exp \int dx\Bigl[ \bar\Psi (i\dsla -e\Asla )\Psi -\frac{1}{4}
F_{\mu\nu}F^{\mu\nu}\Bigr]  
\eeq
($k$ \ldots instanton number).
Therefore, one needs scalar VEVs $\langle S(x_1)\ldots S(x_n)\rangle_0$ of the
massless model, where $S=\bar\Psi \Psi$, $S_\pm =\frac{1}{2}\bar\Psi (1\pm \gaf
)\Psi$. Chiral VEVs $\langle S_{H_1}\ldots S_{H_n}\rangle_0$, $H_i =\pm$, are
especially easily computed, as only a definite instanton sector contributes
(see e.g. \cite{MSSM,GBOUND,Zah}),
\beq
\langle S_{H_1}(x_1)\cdots S_{H_n}(x_n)\rangle_0 = e^{ik\theta} 
\Bigl( \frac{\Sigma}{2}\Bigr)^n \exp
\Bigl[ \sum_{i<j}(-)^{\sigma_i \sigma_j}4\pi D_{\mu_0} (x_i -x_j)\Bigr] 
\eeq
\bdi
k=\sum_{i=1}^n \sigma_i =n_+ -n_-
\edi
where $\sigma_i =\pm 1$ for $H_i =\pm $, $D_{\mu_0}$ is the massive scalar
propagator of the Schwinger boson ($\mu_0^2 =\frac{e^2}{\pi}$) 
and $\Sigma$ is the fermion condensate of the massless model.

The Schwinger boson $\Phi$ is related to the vector current, $J_\mu
=\frac{1}{\sqrt{\pi}}\epsilon_{\mu\nu}\partial^\nu \Phi$, and, therefore,
the $S$ and $\Phi$ VEVs, which we need for the perturbative calculation of
massive VEVs, are related to the vector and scalar current VEVs of the
massless model. Explicitly the $S$ and $\Phi$ VEVs may be computed from the
generating functional (which is at the same time a VEV for $n$ chiral currents)
\bdi
\langle S_{H_1}(x_1)\cdots S_{H_n}(x_n)\rangle_0 [\lambda]= e^{ik\theta} 
\Bigl( \frac{\Sigma}{2}\Bigr)^n \exp
\Bigl[ \sum_{i<j}(-)^{\sigma_i \sigma_j}4\pi D_{\mu_0} (x_i -x_j)\Bigr] \cdot 
\edi
\beq
\cdot \exp\Bigl[-\int dy_1 dy_2 \lambda (y_1) D_{\mu_0}(y_1 -y_2)\lambda (y_2)
+2i\sqrt{\pi}\sum_{l=1}^n (-)^{\sigma_l} \int dy\lambda (y)D_{\mu_0}(y-x_l)
\Bigr] 
\eeq 
(see \cite{Gatt2} for an explicit computation),
where $\lambda$ is the external source for the Schwinger boson $\Phi$. 
Observe the $(-)^{\sigma_l}$ in the last term of the exponent. As a
consequence, whenever an {\em odd} number of external $\Phi$ lines meets at a
point $x_i$, 
the corresponding $S_- (x_i)$ acquires a $-$, i.e. instead of a $S=S_+
+S_-$ vertex there is a $P=S_+ -S_-$ vertex.

From equs. (2), (3) one finds that exponentials $\exp \pm 4\pi D_{\mu_0} (x_i
-x_j)$ are running from any vertex $x_i$ to any other vertex $x_j$; however, in
order to get an IR finite perturbation theory, one has to expand these
exponentials into the functions
\beq
E_\pm (x)=e^{\pm 4\pi D_{\mu_0}(x)} -1,
\eeq
(or their Fourier transforms $\wt E_\pm (p)$ for momentum space Feynman rules).
This expansion is analogous to the cluster expansion of statistical physics.

The mass perturbation theory may be summarized by the following Feynman rules

$$\psannotate{\psboxscaled{800}{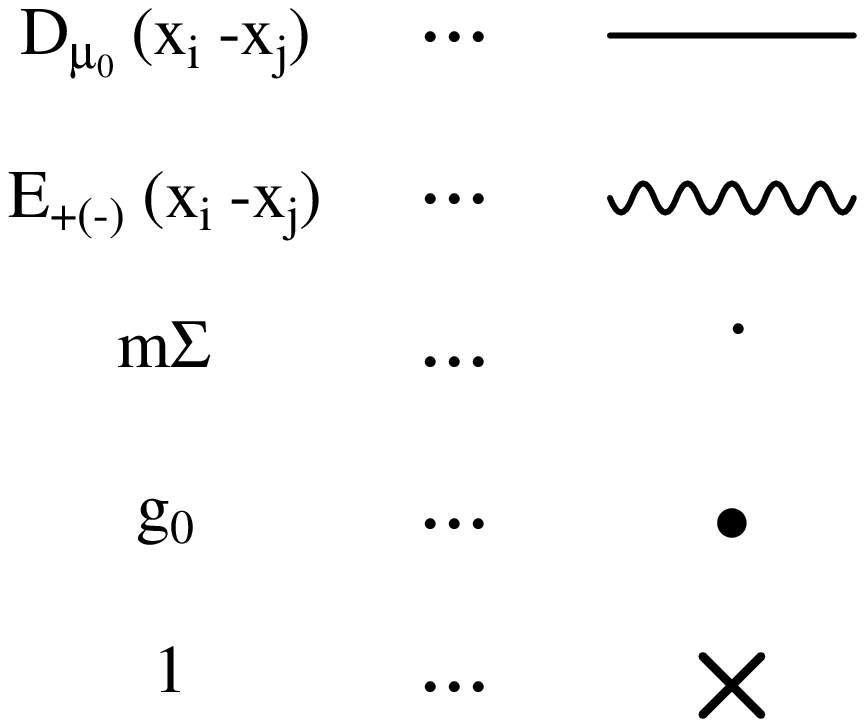}}{\fillinggrid
\at(5.9\pscm;-0.5\pscm){Fig. 1}}$$

where $m\Sigma$ is the bare coupling constant of mass perturbation theory and
$g$ the renormalized one (including all tadpole-like corrections, see e.g.
\cite{GBOUND}). Strictly speaking, we added the $S_+$ and $S_-$ contributions
of each vertex, therefore the rules, as they stand, are true only for the
special case $\theta=0$. The general $\theta$ case we will discuss in a
subsequent section, where we will find that the graphical notation of Fig. 1
may be used, but the meaning of the individual graphs is slightly changed. 

Using this mass perturbation theory the following features could be shown to
hold in \cite{BOUND,GBOUND}:
\begin{enumerate}
\item The mass corrections for the Schwinger boson (computable from $\langle
\Phi (x_1)\Phi (x_2)\rangle_m$, \cite{SMASS}) 
also occur as corrections to the internal boson
lines of the exponentials $E_\pm (x)$. Therefore, the bare Schwinger mass
$\mu_0$ may be replaced by the renormalized one, $\mu$, in all internal
propagators.
\item In the perturbation series of Fig. 1 there occur, 
among other terms, strings of
$E_\pm$, $E_\pm (x_1 -x_2)\cdot E_\pm (x_2 -x_3)\ldots $ 
When the two-boson part
$\frac{16\pi^2}{2!}D_\mu^2 (x)$ is separated in the expanded exponential 
$E_\pm (x)$, (4), these two-boson blobs may be resummed in momentum space,

$$\psannotate{\psboxscaled{800}{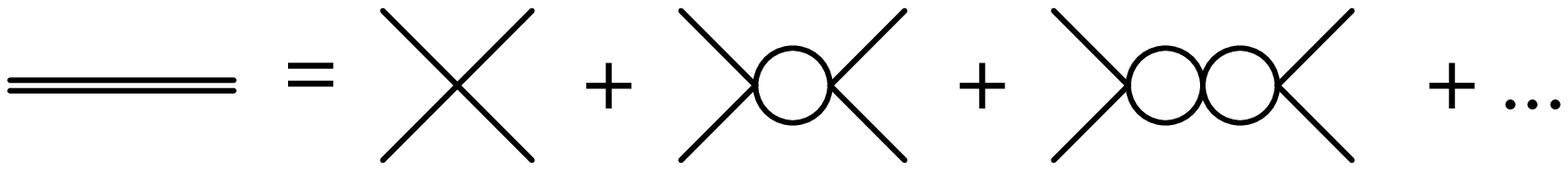}}{\fillinggrid
\at(8\pscm;-0.5\pscm){Fig. 2}}$$

via a geometric series formula and give the leading contribution to the
two-boson bound state mass pole (with mass $M_2$). Higher $n$-boson blobs give
only higher order corrections to the mass, because the two-boson blob is near
its threshold singularity at $M_2$ (see \cite{BOUND}).

Explicitly one finds the mass pole ($s=-p^2 >0$, remember our Euclidean
conventions)
\beq
\frac{1}{1-\frac{16\pi^2}{2!}m\Sigma\cos\theta\frac{1}{\pi
s}\frac{1}{\sqrt{\frac{4\mu^2}{s}-1}}\arctan\frac{1}{\sqrt{\frac{4\mu^2}{s}
-1}}} 
\eeq
with the pole mass
\beq
M_2^2 =4\mu^2 -\Delta_2 \quad ,\quad \Delta_2 =\frac{4\pi^4
(m\Sigma\cos\theta)^2}{\mu^2}
\eeq
and residue at the pole
\beq
R_2 =\frac{2\Delta_2}{m\Sigma\cos\theta}=\frac{8\pi^4
m\Sigma\cos\theta}{\mu^2}.
\eeq
\end{enumerate}

Analogously, we want to show in this paper that
\begin{enumerate}
\item there is a resummation that leads to a $\mu$-$M_2$-blob

$$\psannotate{\psboxscaled{800}{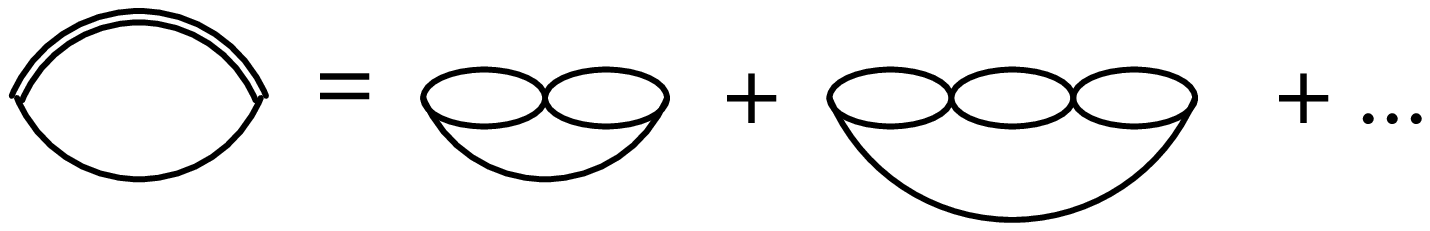}}{\fillinggrid
\at(5.9\pscm;-0.5\pscm){Fig. 3}}$$

(this is quite obvious from the Feynman rules, Fig. 1 and from the fact that
the propagators $E_\pm$ may connect {\em all} vertices);
\item two- and three- boson blobs and the above-mentioned $\mu$-$M_2$-blob (Fig.
3) occur in such combinations that they may again be resummed via the geometric
series formula to result in a propagator
\beq
\frac{1}{1-m\Sigma\cos\theta f(p)}
\eeq
and $f(p)$ is the sum of the three above-mentioned blobs,

$$\psannotate{\psboxscaled{800}{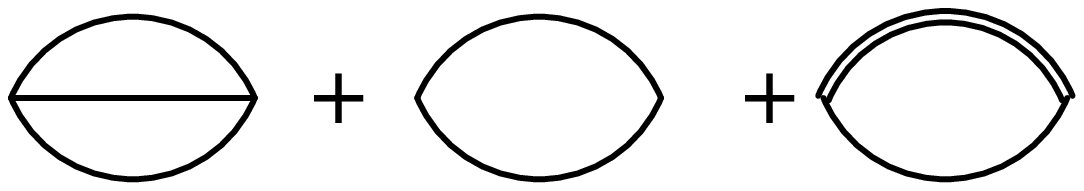}}{\fillinggrid
\at(5\pscm;-0.5\pscm){Fig. 4}}$$

where the three-boson blob will give rise to the three-boson bound-state
mass pole $M_3$, whereas the
$2\mu$- and $\mu$-$M_2$-blobs have imaginary parts that will give the partial
decay widths for the decays of the three-boson bound state $M_3$ into $2\mu$
and $\mu +M_2$, respectively. All other types of blobs that we ignored are
unimportant for these features.

\end{enumerate}

Actually, things are a little bit more complicated. As mentioned above, up to
now the Feynman rules of Fig. 1 are true only for the special case $\theta =0$.
But this restriction is too strong. E.g. the decay $M_3 \ra 2\mu$ is not
possible in that case because of parity conservation. So let us discuss the
general $\theta$ Feynman rules as a next step.

\section{Feynman rules for general $\theta$}

In the mass perturbation theory each vertex is $mS=m(S_+ +S_-)$. For general
$\theta$ the two chiral components $S_+$ and $S_-$ couple differently, with
(renormalized) coupling $g_\theta =m\frac{\Sigma}{2}e^{i\theta} +o(m^2)$ for
$S_+$ and $g^*_\theta =m\frac{\Sigma}{2}e^{-i\theta} +o(m^2)$ for $S_-$. 
As a consequence, the Feynman rules of Fig. 1 acquire a matrix structure. More
precisely, the wavy line (the $E_\pm$ propagator) turns into a matrix,
\beq
{\cal E}(p) =\left( \begin{array}{cc}\wt E_+ (p) & \wt E_- (p) \\ \wt E_- (p)
 & \wt E_+ (p) \end{array} \right)
\eeq
whereas the vertex $g$ turns into an $n$-th rank tensor ${\cal G}$ when $n$
wavy lines ${\cal E}$ meet at this vertex. Only two components of this tensor
are nonzero, namely ${\cal G}_{++\cdots +} =g_\theta$ and ${\cal G}_{--\cdots
-} =g_\theta^*$. E.g. when two lines meet at one vertex, then ${\cal G}$ is the
matrix
\beq
{\cal G} =\left( \begin{array}{cc} g_\theta & 0 \\ 0
 & g_\theta^* \end{array} \right) .
\eeq
External bosons $\Phi$ that meet at a vertex are represented by $P$ ($S$), when
an odd (even) number of bosons meets at the vertex, where
\beq
P = {1 \choose -1} \quad ,\quad S={1 \choose 1} .
\eeq
Among other Feynman graphs, there exist the following ones (that may be
resummed), where we amputate external bosons,

$$\psannotate{\psboxscaled{800}{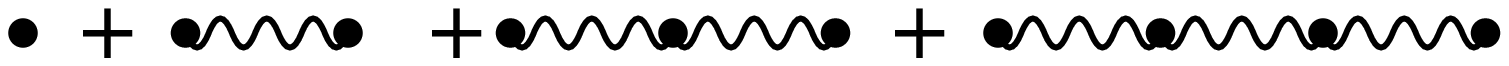}}{\fillinggrid
\at(7\pscm;-0.5\pscm){Fig. 5}}$$

or
\beqa
{\cal G}\Pi (p) &:=& {\cal G} +{\cal G}{\cal E}(p){\cal G}+{\cal G}{\cal E}(p)
{\cal G}{\cal E}(p){\cal G} +\ldots \no \\
&=& {\cal G}({\rm\bf 1} +{\cal E}(p){\cal G}\Pi (p))
\eeqa
with the solution 
\beq
\Pi (p) =\frac{1}{N(p)} \left( \begin{array}{cc} 1-g_\theta^*\wt E_+ (p)
 & g_\theta^* \wt E_- (p) \\ g_\theta \wt E_- (p)
 & 1- g_\theta \wt E_+ (p) \end{array} \right)
\eeq
where
\beq
N(p) =\det ({\rm\bf 1}-{\cal G}{\cal E}(p))= 
1-(g_\theta +g_\theta^* )\wt E_+ (p)
+g_\theta g^*_\theta (\wt E_+^2 (p) -\wt E^2_- (p)).
\eeq
So we succeeded in inverting the propagator $\Pi (p)$, analogously to our above
discussion. 

For the determination of the mass pole and decay widths only the denominator
$N(p)$ of (13) will be important. In the denominator $N(p)$ we find all the
$n$-boson propagators
\beq
d_n (p):= \frac{(4\pi)^n}{n!}\wt{D^n}(p).
\eeq
But we want to include the $\mu ,M_2$ two-boson loop
\beq
d_{1,1}(p):= \int \frac{d^2 q}{(2\pi)^2}\frac{8\pi^4 m\Sigma\cos\theta}{\mu^2
(q^2 +M_2^2 )}\frac{4\pi}{(p-q)^2 +\mu^2} ,
\eeq
too. In principle, its existence is obvious from the Feynman rules of Fig. 1.
But how does it fit into our matrix scheme? The answer is that it is parity odd
and may be treated just like any odd $n$-boson propagator (e.g. like $d_3$), as
may be checked by expanding the perturbation series into the individual
$n$-boson propagators and including the $\mu$-$M_2$ two-boson propagator term by
term. So we just include it into $E_\pm$ with a $\pm$ sign as a higher order
contribution that we need. 

Inserting the $n$-boson propagators $d_n$, we may rewrite the denominator
$N(p)$ like
\bdi
N(p)\simeq
1-(g_\theta +g^*_\theta )(d_1 +d_2 +d_{1,1}+d_3 +\ldots )+
\edi
\beq
4g_\theta g^*_\theta
\Bigl( d_1 (d_2 +d_4 +\ldots ) +d_{1,1}(d_2 +d_4 +\ldots )+d_3 (d_2 +d_4
+\ldots ) +\ldots \Bigr)
\eeq
where we included $d_{1,1}$, as stated above.

Now suppose we are at the three-boson bound-state mass $M_3$. Then the real
part of (17) vanishes by definition, and the leading order contribution (of
order $\frac{1}{g_\theta +g^*_\theta}$) stems from the $d_3 (s=-p^2)$ function
that is near its threshold singularity,
\beq
(g_\theta +g^*_\theta )d_3 (s=M_3^2 )=1+o(m^2).
\eeq
In addition, all functions $d_i$ that are above their respective mass
thresholds will have imaginary parts at $M_3^2$, and result thereby in partial
decay widths. This is obviously so for $d_2$, and it is also true for $d_{1,1}$
provided $M_3 >\mu +M_2$ (actually this is the case as we will see in the
computations). Therefore, at $M_3$ (17) may be rewritten like
\bdi
-i(g_\theta +g^*_\theta )\Bigl( {\rm Im \,}d_2 (M_3^2) +{\rm Im \,}
d_{1,1}(M_3^2)\Bigr) +4ig_\theta
g^*_\theta d_3 (M_3^2) {\rm Im \,} d_2 (M_3^2) +o(m^2)
\edi
\beq
=-im\Sigma\cos\theta \Bigl({\rm Im \,} d_2 (M_3^2) +{\rm Im \,}
d_{1,1}(M_3^2)\Bigr)
+i\frac{m\Sigma}{\cos\theta}{\rm Im \,}d_2 (M_3^2) +o(m^2)
\eeq
and we find that parity forbidden decay widths acquire a factor $(\cos\theta -
\frac{1}{\cos\theta})$, whereas parity allowed decays acquire a $\cos\theta$
factor.

So we have, indeed, succeeded in rewriting the mass pole equation, as indicated
in Fig. 4, and may start the actual computations.

\section{Bound state mass and decay width computations}

For the computation of the 
three-boson bound-state mass $M_3$ we need the three-boson propagator $d_3$ 
and find, in lowest order (see (18))
\beq
1=\frac{1}{3!}m\Sigma \cos \theta \cdot 64\pi^3 \widetilde{D_\mu^3 } (p)
\eeq
or, after a rescaling $p\ra \frac{p}{\mu}$ to dimensionless momenta
\beq
1=\frac{64\pi^3}{6}\frac{m\Sigma}{\mu^2}\cos\theta
\, \widetilde{D_\mu^3 } (p).
\eeq
$\widetilde{D_\mu^3 } (p)$ is given by the graph (where we introduce positive 
squared momentum $s =-p^2 > 0$)

$$\psannotate{\psboxscaled{1000}{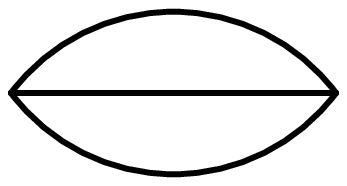}}{\fillinggrid
\at(1.2\pscm;-0.5\pscm){Fig. 6}}$$

\bdi
\widetilde{D_\mu^3 } (p) =
-\int \frac{d^2 q_1 d^2 q_2}{(2\pi)^4}\frac{1}{(p+q_1 +q_2)^2 +1}
\frac{1}{q_1^2 +1}\frac{1}{q_2^2 +1}=
\edi
\bdi
-2\int_0^1 dx\int_0^x dy\int  \frac{d^2 q_1 d^2 q_2}{(2\pi)^4}
\frac{1}{\Big[ q_1^2 +1+(q_2^2 -q_1^2)x + ((p+q_1 +q_2 )^2 -q_2^2)y\Bigr]^3}=
\edi
\bdi
\int \frac{dx}{(4\pi)^2}\int_0^x \frac{dy}{s (xy-x^2 y-y^2 +xy^2 )-x+x^2 -xy 
+y^2} =
\edi
\bdi
\int \frac{dx}{8\pi^2 (1-s (1-x))}\int_0^{\frac{x}{2}}
\frac{dz}{z^2 +T^2 (s ,x)} =
\edi
\beq
\int_0^1 \frac{dx}{8\pi^2 (1-s (1-x))}\frac{1}{T(s ,x)}
\arctan \frac{x}{2T(s ,x)} ,
\eeq
\beq
T^2 (s ,x)=\frac{x^2 -s x^2 (1-x)+4x(1-x)}{4(s (1-x)-1)} ,
\eeq
where, as usual, we introduced Feynman parameter integrals and performed the
momentum integrations. Further, the first Feynman parameter integral could be
done analytically. 
The numerator of $T^2$ has a double zero at $s =9$:
\beq
9x(x-\frac{2}{3})^2 .
\eeq
This double zero is in the integration range of $x$ and is precisely the 
threshold singularity. Setting
\beq
s=9-\Delta_3
\eeq
in the numerator of $T^2$ in the factor $\frac{1}{T}$, and $s =9$ everywhere
else, where it is safe, one arrives at:
\beq
\frac{1}{12\pi^2}\int_0^1 \frac{dx}{\sqrt{\vert 9x-8\vert }}
\frac{\arctan \frac{\sqrt{x\vert 9x-8\vert }}{3(x-\frac{2}{3})}}{
\sqrt{(x-\frac{2}{3})^2 x+\frac{\Delta_3}{9} x^2 (1-x)}}=:I(\Delta_3 ).
\eeq
The mass-pole equation reads 
\beq
1=\frac{64\pi^3}{6}m\Sigma\cos\theta I(\Delta_3 )
\eeq
and must be evaluated numerically. It gives rise to an extremely tiny mass 
correction $\Delta_3$. For sufficiently small $m$ it is very well 
saturated by
\beq
\Delta_3 (m\Sigma\cos\theta )
\simeq 6.993 \exp (-\frac{0.263}{m\Sigma\cos\theta})
\eeq
and is therefore smaller than polynomial in $m$. (I checked the 
numerical formula (28) for $30 <\frac{1}{m\Sigma\cos\theta}<1000$, 
corresponding to
$10^{-2}<\Delta_3 <10^{-100}$, but I am convinced that it remains true for
even larger $\frac{1}{m\Sigma\cos\theta}$; 
however, there the numerical integration
is quite difficult because of the pole in (26).) 

We conclude that the three-boson bound state mass is nearly entirely given by 
three times the Schwinger boson mass 
(we change back to dimensionfull quantities now),
\beq
M_3^2 = 9\mu^2 -\Delta_3 \quad ,\quad \Delta_3 =6.993\mu^2 \exp
(-0.263\frac{\mu^2}{m\Sigma\cos\theta})
\eeq
or, differently stated, that the binding of three bosons is extremely weak.

Therefore it holds that $M_3 >\mu +M_2$, as indicated above, and, consequently, 
a decay of
$M_3$ into $\mu +M_2$ is possible. This has the consequence that the
three-boson bound state is unstable even for $\theta =0$, contrary to some
earlier conjecture (\cite{Co1}). 

So let us turn to the decay width computation. For this purpose we need the
first Taylor coefficient of the denominator $1-m\Sigma\cos\theta d_3 (s)$ (see
(14)) around the mass pole. Observe that because of formulae (25), (28)
$m\Sigma\cos\theta d_3 (s)$ may be written, in the vicinity of $s=M_3^2$, like
\beq
m\Sigma\cos\theta d_3 (s)\sim \frac{m\Sigma\cos\theta}{0.263}\ln\frac{6.993
\mu^2}{9\mu^2 -s}.
\eeq
Therefore, we find the Taylor coefficient
\beq
c_3 =\frac{m\Sigma\cos\theta}{0.263 \Delta_3}.
\eeq
Generally, a decay width may be inferred from the imaginary part of a
propagator,
\beq
G(p)\sim \frac{{\rm const.}}{s-M^2 -iM\Gamma} ,
\eeq
where $\Gamma$ is the decay width.

In our case we have for $\frac{1}{N(p)}$, equ. (19),
\bdi
\frac{1}{N(p)}
\sim \frac{1}{c_3 (s-M_3^2 )-im\Sigma (\cos\theta -\frac{1}{\cos\theta}) {\rm
Im \,}d_2 (M_3^2) - im\Sigma\cos\theta {\rm Im \,}d_{1,1} (M_3^2)}
\edi
\beq
\simeq \frac{{\rm const.}}{s-M_3^2 -i\frac{m\Sigma}{c_3}\Bigl[ (\cos\theta
-\frac{1}{\cos\theta}){\rm Im \,}d_2 (M_3^2) + \cos\theta {\rm Im
\,}d_{1,1}(M_3^2)\Bigr] } .
\eeq
Next we need the imaginary parts ${\rm Im \,}d_2$, ${\rm Im \,}d_{1,1}$. Both
of them stem from a two-boson blob, so let us write down the general result
(which is standard) ($s=-p^2$)
\beqa
{\rm Im \,}\wt{(D_{M_1}D_{M_2})}(s)&=& {\rm Im \,}\int \frac{d^2 q}{(2\pi)^2}
\frac{-1}{q^2 +M_1^2}\frac{-1}{(p-q)^2 +M_2^2} \no \\
&=& \frac{1}{2w(s,M_1^2 ,M_2^2 )},
\eeqa
\beq
w(x,y,z)=(x^2 +y^2 +z^2 -2xy-2xz-2yz)^\frac{1}{2} .
\eeq
Therefore we can write for (34) (see (7) and (16) for the normalization factors
of $d_2$ and $d_{1,1}$)
\beq
\frac{{\rm const.}}{s-M_3^2 -i\frac{m\Sigma}{c_3}(\cos\theta
-\frac{1}{\cos\theta})\frac{4\pi^2}{w(M_3^2 ,\mu^2 ,\mu^2 )}
-i\frac{(m\Sigma\cos\theta )^2}{c_3}\frac{16\pi^5}{\mu^2 w(M_3^2 ,\mu^2 ,M_2^2
)}}.
\eeq
There seems to be something wrong with the sign of the parity forbidden partial
decay width $\Gamma_{M_3 \ra 2\mu}$ (the $d_2$ term). Actually the sign is o.k.
and the problem is a remnant of the Euclidean conventions that are implicit in
the whole computation (see e.g. \cite{ABH,MSSM}). In these conventions $\theta$
is imaginary and therefore $(\cos\theta -\frac{1}{\cos\theta}) \ge 0$. 
Of course,
this is not a reasonable convention for a final result. When performing the
whole computation in Minkowski space and for real $\theta$, roughly speaking,
the roles of $E_+$ and $E_-$ are exchanged in (14). This gives an additional
relative sign between parity even and odd $n$-boson propagators and, therefore,
changes the factor of $d_2$ to $(\frac{1}{\cos\theta} -\cos\theta )$, which is
$\ge 0$ for real $\theta$.

With this remark in mind, and expressing the final results for real $\theta$,
we find, by using the approximations
\beq
w(M_3^2 ,\mu^2 ,\mu^2 )\simeq w(9\mu^2 ,\mu^2 ,\mu^2 )=3\sqrt{5}\mu^2
\eeq
\beq
w(M_3^2 ,M_2^2 ,\mu^2 )\simeq w(9\mu^2 ,M_2^2 ,\mu^2 )=2\sqrt{3}\mu
\sqrt{\Delta_3} +o(m^2),
\eeq
the following results:
\beqa
\Gamma_{M_3 \ra 2\mu}&=&
0.263\frac{4\pi^2\Delta_3}{9\sqrt{5}\mu}(\frac{1}{\cos^2 \theta}-1) \no \\
&\simeq & 3.608 \mu (\frac{1}{\cos^2 \theta}-1)\exp
(-0.929\frac{\mu}{m\cos\theta})
\eeqa
and
\beqa
\Gamma_{M_3 \ra \mu +M_2}&=& 0.263\frac{4\pi^3 \Delta_3}{3\sqrt{3}\mu} \no \\
&\simeq & 43.9 \mu \exp (-0.929\frac{\mu}{m\cos\theta})
\eeqa
where we inserted the numerical value $\Sigma =\frac{e^\gamma \mu}{2\pi} =0.283
\mu$. 

The ratio of the two partial decay widths does not depend on the approximations
that were used for the $M_3$ computation,
\beq
\frac{\Gamma_{M_3 \ra 2\mu}}{\Gamma_{M_3 \ra \mu +M_2}}= \frac{\frac{1}{\cos^2
\theta}-1}{\sqrt{15}\pi}.
\eeq
Contrary to a naive expectation it holds that $\Gamma_{M_3 \ra \mu +M_2}>
\Gamma_{M_3 \ra 2\mu}$, although $\mu +M_2 \sim M_3$. This is not surprizing in
two dimensions, because there the phase space "volume" does not grow with
increasing momentum.

\section{Summary}

So we have succeded in computing the three-boson bound state mass and partial
decay widths via (resummed) mass perturbation theory. The computations may be
generalized and lead to the following physical picture: there are two stable
particles in the theory, namely the fundamental Schwinger boson $\mu$ and the
two-boson bound state $M_2$. Higher bound states are unstable (resonances) and
may decay into all combinations of $\mu$ and $M_2$ final particles that are
allowed kinematically.

There exists also another way for the computation of partial decay widths,
namely the usual perturbative method of putting all the initial and final
states on their mass shells, inserting the squared transition matrix element
and summing over all kinematically possible final states. In our theory this
needs the construction of the exact higher $n$-point functions (here the
three-point function) as a prerequisite, but once this is done it is easy to
show that the decay widths (40), (41) just correspond to a first order
perturbative computation (there the nontrivial numerical factors in (40), (41)
are caused by the
proper normalizations of the initial and final states, which in our computation
correspond to the residues of the mass poles).

Observe that the decay widths (40), (41) are related to the $M_3$ 
binding energy,
$\Gamma_{M_3} \sim \Delta_3$. This is a very reasonable result that enables us
to interpret the $M_3$ bound state (and, a fortiori, higher bound states) as a
resonance. Indeed, suppose that the propagator (13) belongs to the matrix
element of a scattering process (to be discussed elsewhere \cite{SCAT}). The
denominator of (13) has zero real part at $s=M_3^2$ and infinite real part at
the real production threshold $s=9\mu^2 =M_3^2 +\Delta_3$, therefore it will
give a local maximum of the scattering cross section 
(resonance) near $M_3^2$, and a local minimum at $9\mu^2$,
and the resonance width {\em must} be related to the binding energy,
$\Gamma_{M_3} \sim \Delta_3$.

\section*{Acknowledgement}

The author thanks the members of the Institute of Theoretical Physics of the
Friedrich-Schiller-Universit\"at Jena, where this work was done, for their
hospitality. Further thanks are due to A. Wipf and 
J. Pawlowski for helpful discussions.

This work was supported by a research stipendium of the Vienna University.


\begin{thebibliography}{999999}
\bibitem{CJS}
S. Coleman, R. Jackiw, L. Susskind, Ann. Phys. {\em 93} (1975) 267
\bibitem{Co1}
S. Coleman, Ann. Phys. {\em 101} (1976) 239
\bibitem{AAR}
E. Abdalla, M. Abdalla, K. D. Rothe, "2 dimensional Quantum Field Theory",
World Scientific, Singapore, 1991
\bibitem{GKMS}
D. J. Gross, I. R. Klebanov, A. V. Matytsin, A. V. Smilga, Nucl. Phys. 
{\em B461} (1996) 109, HEP-TH 9511104
\bibitem{CONF}
C. Adam, preprint FSUJ TPI 13/96, HEP-TH 9609155
\bibitem{BOUND}
C. Adam, preprint PM 96/01, HEP-PH 9601227, to be published in Z. Phys. {\em C}
\bibitem{GBOUND}
C. Adam, Phys. Lett. {\em B 382} (1996) 111; HEP-TH 9602175
\bibitem{SCAT}
C. Adam, "Scattering processes in the massive Schwinger model", 
to be published
\bibitem{FS1}
J. Fr\"ohlich, E. Seiler, Helv. Phys. Acta {\em 49} (1976) 889
\bibitem{MSSM}
C. Adam, Phys. Lett. {\em B 363} (1995) 79; HEP-PH 9507279
\bibitem{SMASS}
C. Adam, Phys. Lett. {\em B 382} (1996) 383; HEP-PH 9507331
\bibitem{Sc1}
J. Schwinger, Phys. Rev. {\em 128} (1962) 2425
\bibitem{LS1}
J. Lowenstein, J. Swieca, Ann. Phys. {\em 68} (1971) 172
\bibitem{Jay}
C. Jayewardena, Helv. Phys. Acta {\em 61} (1988) 636
\bibitem{SW1}
I. Sachs, A. Wipf, Helv. Phys. Acta {\em 65} (1992) 653
\bibitem{IP1}
N. P. Ilieva, V. N. Pervushin, Sov. J. Part. Nucl. {\em 22} (1991) 275
\bibitem{ABH}
C. Adam, R. A. Bertlmann, P. Hofer, Riv. Nuovo Cim. {\em 16}, No
8 (1993)
\bibitem{Adam}
C. Adam, Z. Phys. {\em C63} (1994) 169
\bibitem{Zah}
J. Steele, A. Subramanian, I. Zahed, HEP-TH 9503220
\bibitem{Gatt2}
C. Gattringer, HEP-TH 9503137, 9603010
\end{thebibliography}
\end{document}